\documentclass[aps,pre,groupedaddress,amsmath,runinaddress,showpacs,twocolumn]{revtex4}
\usepackage{graphicx}
\usepackage{subfigure}
\usepackage{amssymb}
\renewcommand{\vec}[1]{\mathbf{#1}}

\begin{document}

\title{Fluid Modes of a Spherically Confined Yukawa Plasma}


\author{H. K\"{a}hlert}
\author{M. Bonitz}
\affiliation{Institut f\"{u}r Theoretische Physik und Astrophysik, Christian-Albrechts Universit\"{a}t zu Kiel, 24098 Kiel, Germany}



\date{\today}

\begin{abstract}
The normal modes of a three-dimensional Yukawa plasma in an isotropic, harmonic confinement are investigated by solving the linearized cold fluid equations. The eigenmodes are found analytically and expressed in terms of hypergeometric functions. It is found that the mode frequencies solely depend on the dimensionless plasma parameter $\xi=\kappa R$, where $R$ is the plasma radius and $\kappa$ the inverse screening length. The eigenfrequencies increase monotonically with $\xi$ and saturate in the limit $\xi\to\infty$. Compared with the results in the Coulomb limit~[D. H. E. Dubin, Phys. Rev. Lett. \textbf{66}, 2076 (1991)], we find a new class of modes characterized by the number $n$ which determines the number of radial nodes in the perturbed potential. These modes originate from the degenerate bulk modes of the Coulomb system. Analytical formulas for the eigenfrequencies are derived for limiting cases.
\end{abstract}


\maketitle

\section{Introduction}
The interaction between charged particles is known to be strongly affected by a background plasma. Examples include dusty plasmas, where the screening of the dust-dust interaction is mainly determined by ions~\cite{bonitz2010rev,morfill2009rev,piel2008rev}, and dense two-component plasmas~\cite{gericke2010,patrick2010,bonitz2005,kremp2005}, where the ions are screened by weakly coupled electrons. These plasmas are expected to occur in the interior of giant planets and white dwarf stars. While in the former case the degree of screening is determined by the ion Debye length, the screening length in the latter is the Thomas-Fermi length, owing to the degeneracy of the electrons. The Yukawa one-component plasma model is often used to describe the heavy plasma component while the light component determines the screening length. Its static properties and collective excitations have been studied in several publications, e.g.~\cite{hamaguchi1997,kalman2000,donko2008}.

In many situations the plasma is neither homogeneous nor macroscopic. Recently, it was shown~\cite{Christian2006,Christian2007} that the density of a three-dimensional dusty plasma, where gravity is balanced by a thermophoretic force~\cite{arp2004}, is not homogeneous. The reason is the screened dust-dust interaction which produces an inhomogeneous density profile in a harmonic confinement. This is different from experiments with confined ions~\cite{mortensen2006} where the interaction is Coulombic and the mean density is constant. In astrophysical plasmas the confinement is provided by gravity and may also influence the plasma properties.

Previous continuum theories~\cite{Christian2006,Christian2007} for Yukawa plasmas were limited to static properties. Here we extend these results to a time-dependent theory and investigate the normal modes of a Yukawa plasma in a spherical, harmonic confinement~\cite{Bonitz2006,kaehlert2010}. This model is appropriate for the experiments of~\cite{arp2004,block2008}, for which the normal modes of rather small dust crystals have recently been measured~\cite{ivanov3D}. On the one hand, a fluid approach is expected to be accurate for long wavelength modes in a weakly coupled plasma. On the other hand, the agreement of theoretical predictions~\cite{dubin_prl91} with experiments~\cite{drewsen2010,bollinger1993} and simulations~\cite{dubin_md96} for confined ions turns out to be surprisingly good even in the strongly coupled phase. An analogous result for a confined one-component plasma with a screened interaction is still missing. Open questions are the influence of screening on the normal modes and the eigenfrequencies. Compared with the Lagrangian description of Ref.~\cite{sheridan2006} for the breathing mode, the present approach makes no assumption about the particular mode form. Besides dusty plasmas and compact star interiors, we expect our results to be relevant for other systems as well, when screening and confinement are not negligible.

This paper is organized as follows. The fluid equations are introduced and linearized in Sec.~\ref{sec:fluideq}. In Sec.~\ref{sec:solution} we explicitly consider an isotropic harmonic confinement. The density profile is reviewed and used to calculate the ground state potential and energy. Further, the linearized Poisson equation is solved and the eigenfrequency spectrum is derived. The normal modes are discussed in detail. We conclude with a discussion of the theory and an outlook on future work in Sec.~\ref{sec:conclusion}.

\section{Fluid description}\label{sec:fluideq}
\subsection{Basic equations}
The fluid equations for a spatially confined one-component plasma read
\begin{subequations}\label{eqn:fluid}
\begin{align}
\frac{\partial n}{\partial t} +\nabla \cdot (n \vec v)&=0,\\
m n \left[\frac{\partial \vec v}{\partial t}+(\vec v\cdot \nabla) \vec v\right] &= -n\nabla U -\nabla\cdot \vec P -m n \nu \vec v,\label{eqn:eomfluid}
\end{align}
\end{subequations}
where $U(\vec r)=V(\vec r)+q\phi(\vec r)$ denotes the sum of the confinement potential $V(\vec r)$ and the potential $\phi(\vec r,t)$ induced by the particles. In the first (continuity) equation $n(\vec r,t)$ is the particle density and $\vec v(\vec r,t)$ their mean velocity. The second equation is the momentum equation, where $m$ denotes the particle mass, $q$ their charge and $\vec P(\vec r,t)$ the pressure tensor. A damping term with friction coefficient $\nu$ is included to account for collisions with neutral particles.

The fluid equations are complemented by Poisson's equation for the (induced) potential $\phi$,
\begin{equation}\label{eqn:poisson}
 \left(\Delta - \kappa^2\right) \phi = -4 \pi q n,
\end{equation}
where the screening of the interaction between the heavy particles by a polarizable background medium (light charged components) is explicitly taken into account. The range of the interaction is determined by the inverse of the screening parameter $\kappa$.

\subsection{Linearization}
A small perturbation of the plasma equilibrium is well described by linear response theory. This could be caused by an external perturbation (e.g. laser manipulation of particles in a dusty plasma) or by thermal effects. We are interested in strongly coupled plasmas and hence can neglect the pressure term (cold fluid limit). Eqs.~(\ref{eqn:fluid},\ref{eqn:poisson}) are then linearized according to $n(\vec r,t)\simeq n_0(\vec r)+n_1(\vec r,t),\; \vec v(\vec r,t)\simeq\vec v_1(\vec r,t),\; \phi(\vec r,t)\simeq\phi_0(\vec r)+\phi_1(\vec r,t)$. Products of first order terms are assumed negligible.

The classical equilibrium density profile $n_0(\vec r)$ and the associated potential $\phi_0(\vec r)$ are determined from the zero order terms of Eqs.~(\ref{eqn:eomfluid}, \ref{eqn:poisson}),
\begin{subequations}\label{eqn:groundstate}
\begin{align}
q\nabla \phi_0&=-\nabla V,\\
\left(\Delta -\kappa^2\right) \phi_0&=-4\pi q n_0,\label{eqn:potential0}
\end{align}
\end{subequations}
which describe local force equilibrium and are equivalent to the energy minimization in~\cite{Christian2006} (we neglect the finite size factor $(N-1)/N$, where $N$ is the particle number).

First order quantities are determined by
\begin{subequations}\label{eqn:perturbed}
\begin{align}
\frac{\partial  n_1}{\partial t}+\nabla \cdot (n_0 \vec v_1) &=0,\\
m\frac{\partial \vec v_1}{\partial t}+m\nu \vec v_1 &= -q\nabla \phi_1,\\
\left(\Delta -\kappa^2\right) \phi_1&=-4\pi q n_1.\label{eqn:pertpotential}
\end{align}
\end{subequations}
Looking for normal mode solutions with a time dependence $e^{-i\omega t}$, e.g. $\phi_1(\vec r,t)=\hat\phi_1(\vec r) e^{-i\omega t}$, we obtain
\begin{subequations}\label{eqn:firstorder}
\begin{align}
i\omega \hat{n}_1 &={\nabla\cdot (n_0 \hat{\vec v}_1)},\\
m(\omega+i\nu)\hat{\vec v}_1&= -{iq}{\nabla \hat{\phi}_1}.\label{eqn:vel1}
\end{align}
\end{subequations}
Using Eqs.~(\ref{eqn:firstorder}) we can rewrite~(\ref{eqn:pertpotential}) as
\begin{equation}\label{eqn:poisson_diel}
 \nabla \cdot \left[ \epsilon(\vec r,\omega) \nabla \hat{\phi}_1\right]=\kappa^2 \hat{\phi}_1,
\end{equation}
where the plasma dielectric function is given by
\begin{equation}\label{eqn:phi1}
 \epsilon(\vec r,\omega)=1-\frac{\omega_{p}^2(\vec r)}{\omega(\omega +i\nu)}
\end{equation}
and 
\begin{align}\label{eqn:plasmafreq}
\omega_{p}(\vec r)= \sqrt{4\pi q^2 n_0(\vec r)/m}
\end{align}
denotes the local plasma frequency. $\hat{n}_1$ and $\hat{\vec v}_1$ have been eliminated in favor of $\hat{\phi}_1$.

Eq.~(\ref{eqn:poisson_diel}) is a self-contained equation for $\hat{\phi}_1$ and will be solved in the following section for a special case. Having found its solution, $\hat{n}_1$ and $\hat{\vec v}_1$ follow from Eqs.~(\ref{eqn:firstorder}).

\section{Solution for harmonic confinement}\label{sec:solution}

\subsection{Ground state}
So far our results are valid for arbitrary confinement. In order to make further progress let us now explicitly consider an isotropic harmonic confinement $V(r)=m\omega_0^2 r^2/2$. The ground state density $n_0(r)$ [cf. Eqs.~(\ref{eqn:groundstate})] is given by~\cite{Christian2006}
\begin{equation}\label{eqn:density0}
 n_0(r)=\frac{3}{4\pi a^3}\left(1+\frac{\xi^2}{6}\frac{3+\xi}{1+\xi}-\frac{\kappa^2 r^2}{6}\right)\Theta(R-r),
\end{equation}
where $a=(q^2/m\omega_0^2)^{1/3}$ is the Wigner-Seitz radius in the Coulomb limit, $\kappa=0$. The normalized cluster radius is denoted by $\xi=\kappa R$. For Coulomb interaction the density is constant, $\xi\equiv 0$, and $R(\kappa=0)\equiv R_C=aN^{1/3}$, while for $\kappa\ne 0$ $n_0(r)$ decreases parabolically towards the boundary. In this case $R(\kappa)=\xi/\kappa$ must be determined from~\cite{Christian2006}
\begin{equation}\label{eqn:radius}
 \xi^6+6\,\xi^5+15 \left[\xi^4+\xi^3 -k_C^3(\xi+1)\right]=0,
\end{equation}
where $k_C=\kappa R_C$ is the inverse screening length normalized by the Coulomb radius. For small $k_C$ the asymptotic solution of Eq.~(\ref{eqn:radius}) is
\begin{align}
  \xi(k_C)\simeq k_C-\frac{2}{15}k_C^3+\frac{1}{9}k_C^4-\frac{1}{25}k_C^5+\dots ,
\end{align}
while for $k_C\gg 1$
\begin{align}\label{eqn:radius_asy}
\xi(k_C)\simeq&\, 15^{1/5} \,k_C^{3/5} - 1+ \frac{1}{15^{2/5}}\, k_C^{-6/5}\nonumber \\
 &-15^{1/5}\frac{3}{25}\,k_C^{-12/5}+\frac{1}{15}\,k_C^{-3}+\dots \, .
\end{align}
The relative error of these approximations is $<10^{-2}$ for $k_C\le 1.26$ and $k_C>1.26$, respectively. 

The case $\xi\gg 1$ is encountered if $k_C=\kappa a\, N^{1/3}\gg 1$, i.e. if either $\kappa a$ and/or $N$ are large. This is why we will refer to $\xi\to\infty$ as the macroscopic/strong screening limit. The plasma has a size of many screening lengths. The opposite case, $\xi\ll 1$, will be referred to as the Coulomb limit, where the screening length is much larger than the plasma radius.

\begin{figure}
\includegraphics[width=0.47\textwidth]{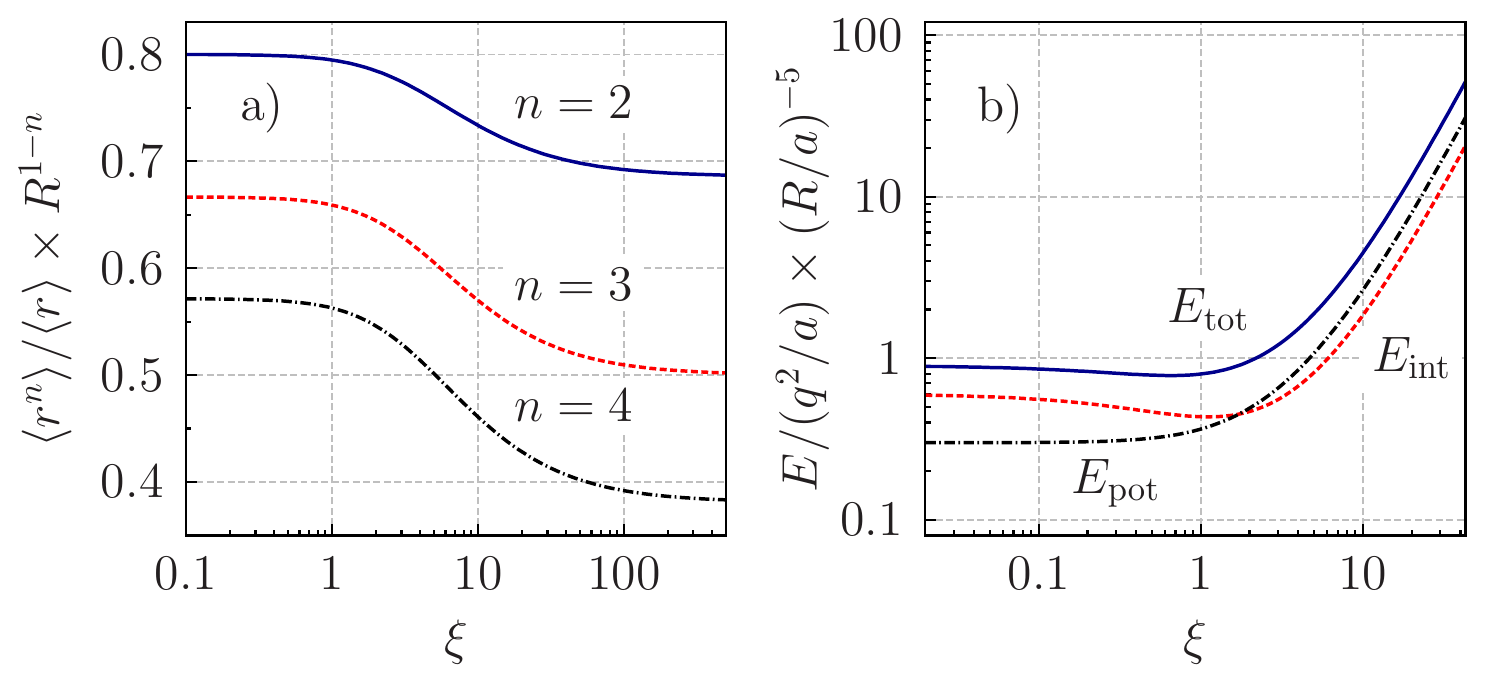}
\caption{Screening dependence of a) ratio of $n$-th to first density moment, and b) ground state energy contributions.}\label{fig:momentsenergy}
\end{figure}

It is straightforward to calculate the moments of the density, which are given by
\begin{align}
 \langle r^n\rangle&=\frac{1}{N}\int r^n \,n_0(r) \, d\vec r\\
&= \frac{R^n}{N}\left(\frac{R}{a}\right)^3\frac{\xi^3+(n+6)\xi^2+3(n+5)(1+\xi)}{(n+5)(n+3)(1+\xi)}.\nonumber
\end{align}
The ratio of two moments [Fig.~\ref{fig:momentsenergy}a] could help determine the unknown parameters $\xi$ and $R$ in experiments, where the particle positions are directly accessible. The moments can easily be calculated since the integral reduces to a sum over all particles as $n(\vec r)=\sum_{i}\delta(\vec r-\vec r_i)$.

The ground state potential $\phi_0(r)$ is determined by Eq.~(\ref{eqn:potential0}) for which the Yukawa potential is the associated Green's function~\cite{Christian2006}. Thus, the solution is given by (details can be found in Appendix~\ref{sec:AppendixA})
\begin{align}\label{eqn:potential0sol}
\phi_0(r)&=q\int  n_0(r')\frac{e^{-\kappa |\vec r-\vec r'|}}{|\vec r-\vec r'|}\,d\vec r' \\
&=\frac{q}{a}\frac{(R/a)^2}{1+\xi}\times
\begin{cases}
\frac{1}{2}\left[3+\xi-(1+\xi)\left[\frac{r}{R}\right]^2\right], & r\le R,\nonumber \\
R \exp(\xi-\kappa r)/r, &r>R.
\end{cases}
\end{align}
Since the confinement is parabolic the potential inside the plasma must decrease correspondingly to ensure force equilibrium. Outside the cloud the potential behaves like that of a point charge placed at the origin. While for Coulomb interaction the (effective) charge is $Q_\text{eff}=Nq$, as expected from Gauss's law, the result for Yukawa interaction  is $Q_\text{eff}=q\,(R/a)^3\,e^{\xi}/[1+\xi]$.

The ground state density and potential can further be used to calculate the total energy in mean-field approximation, $E_\text{tot}=E_\text{pot}+E_\text{int}$, where~\cite{Christian2006}
\[
  E_\text{pot}=\int V(r)n_0(r)d\vec r,\:\: E_\text{int}=\frac{q}{2}\int \phi_0(r) n_0(r)d\vec r.
 \]
Using Eqs.~(\ref{eqn:density0},\ref{eqn:potential0sol}) we find after some algebra,
\begin{align}
  \frac{E_\text{pot}}{q^2/a}&=\frac{(R/a)^5}{10}\left[3+\frac{\xi^2}{7}\frac{8+\xi}{1+\xi}\right],\\
  \frac{E_\text{int}}{q^2/a}&=\frac{(R/a)^5}{210}\left[\frac{126+147\xi+72\xi^2+18\xi^3+2\xi^4}{(1+\xi)^2}\right].\nonumber
 \end{align}
The total energy then reads
\begin{equation*}
 \frac{E_\text{tot}}{q^2/a}=\frac{(R/a)^5}{210}\left[ \frac{189+273\xi+159\xi^2+45\xi^3+5\xi^4}{(1+\xi)^2}\right],
\end{equation*}
see Fig.~\ref{fig:momentsenergy}b. For small $\xi$ the interaction energy yields the dominant contribution to the total energy since the potential is only weakly screened. For large $\xi$ the cluster has a size of several screening lengths and the potential energy dominates. The critical point is at $\xi\approx 1.72$.

For Coulomb interaction ($\xi=0$) the result is $E_\text{tot}=9/10\,N^{5/3}\,q^2/a$, which is the first (mean-field) term in the energy expansion of the shell model (or the energy of the neutralizing background)~\cite{cioslowski2008,Tsuruta1993} . In the opposite limit, $\xi\gg 1$, the asymptote is
\begin{equation}
\frac{E_\text{tot}}{q^2/a}\simeq \left(\frac{R}{a}\right)^5 \frac{\xi^2}{42}= \frac{1}{(\kappa a)^5} \frac{\xi^7}{42}\simeq \frac{15^{7/5}}{42}\frac{N^{7/5}}{(\kappa a)^{4/5}},
\end{equation}
since $\xi\simeq 15^{1/5}\,k_C^{3/5}$ [leading order term in Eq.~(\ref{eqn:radius_asy})]. In the limit $N\gg 1$ we thus have $E_\text{tot}\propto N^{5/3}$ for $\kappa=0$ and $E_\text{tot}\propto\kappa^{-4/5} N^{7/5}$ for finite screening.

\subsection{Normal modes}
Since the ground state density profile terminates in a finite step, cf. Eq.~(\ref{eqn:density0}), one has to solve~(\ref{eqn:poisson_diel}) separately for $r\le R$ and $r>R$. The dielectric function inside and outside the plasma reads
\begin{equation}\label{eqn:dielectric}
 \epsilon(\kappa r,\omega)=
\begin{cases}
 \epsilon_1(\omega)+\frac{\kappa^2 r^2}{2\Omega^2}, &r\le R,\\
1, &r>R,
\end{cases}
\end{equation}
where the constant term is given by
 \begin{align}\label{eqn:epsdef}
\epsilon_1(\omega)&=1-\frac{\omega_p^2(0)}{\omega(\omega+i\nu)}= 1-\frac{\Omega_p^2(0)}{\Omega^2}\nonumber \\
&=1-\frac{1}{\Omega^2}\left(3+\frac{\xi^2}{2}\frac{3+\xi}{1+\xi} \right),
 \end{align}
and $\Omega^2=\omega(\omega+i\nu)/\omega_0^2$. Accordingly we define a normalized plasma frequency $\Omega_p(\kappa r)=\omega_p(r)/\omega_0$. Note that the plasma frequency $\omega_p(r)$ only depends on the product $\kappa r$, see Eqs.~(\ref{eqn:plasmafreq},\ref{eqn:density0}).

Eq.~(\ref{eqn:poisson_diel}) must be supplemented by the boundary conditions
\begin{subequations}\label{eqn:boundarycond}
 \begin{align}
\left.\hat{e}_t\cdot \nabla \hat{\phi}_1^{\text{in}}(\vec r,\omega)\right|_{r=R} &=\left.\hat{e}_t\cdot\nabla \hat{\phi}_1^{\text{out}}(\vec r)\right|_{r=R},\label{eqn:bound1}\\
  \left.\hat{e}_r\cdot\epsilon(\kappa r,\omega) \nabla \hat{\phi}_1^{\text{in}}(\vec r,\omega)\right|_{r=R} &= \left.\hat{e}_r\cdot \nabla \hat{\phi}_1^{\text{out}}(\vec r)\right|_{r=R},\label{eqn:bound2}\\
  \lim_{r\to \infty}\hat{\phi}_1^{\text{out}}(\vec r)&=0.\label{eqn:bound_infinity}
 \end{align}
\end{subequations}
Here $\hat e_t$ and $\hat{e}_r$ are unit vectors in the tangential and radial direction at the surface of the sphere with radius $R$. These are the usual boundary conditions for the tangential component of the electric field $-\nabla \phi_1$ and the radial component of $-\epsilon \nabla \phi_1$.

In order to solve Eq.~(\ref{eqn:phi1}) we use an expansion in spherical harmonics, i.e.
\begin{equation*}
 \hat{\phi}_1(\vec r,\omega) \sim f(r,\omega) Y_{\ell}^m(\theta,\varphi).
\end{equation*}
Since the spherical harmonics are eigenfunctions of the angular part of the Laplacian,
\begin{equation*}
 \Delta Y_{\ell}^m(\theta,\varphi)=-\frac{\ell(\ell+1)}{r^2}Y_{\ell}^m(\theta,\varphi),
\end{equation*}
this leads to the following equation for the radial function $\tilde{f}(x,\omega)$,
\begin{align}\label{eqn:radialx}
 \frac{\partial}{\partial x}\left[\epsilon(x,\omega)x^2 \tilde f'(x,\omega)\right]&\\
-\left[x^2 + {\ell(\ell+1)}\epsilon(x,\omega)\right]\tilde f(x,\omega)&=0,\nonumber
\end{align}
after multiplying by $x^2$. Here we made a change of variables from $r$ to the dimensionless radius $x=\kappa r$ and introduced a new notation $f(r)\to\tilde f(x)$. In the remainder of this section we will separately solve~(\ref{eqn:radialx}) inside and outside the plasma.

Consider first the situation outside the plasma where the dielectric function is just a constant. Here Eq.~(\ref{eqn:radialx}) reduces to
\begin{equation}\label{eqn:radialout}
 x^2 \tilde f''(x)+2x \tilde f'(x)-\left[x^2+\ell(\ell+1)\right]\tilde f(x)=0,
\end{equation}
the solutions of which are modified spherical Bessel functions of the first and second kind, $i_\ell(x)$ and $k_\ell(x)$, respectively~\cite{wolfram}. They are related to the modified Bessel functions by
\[
 i_\ell(x)=\sqrt{\frac{\pi}{2x}}I_{\ell+1/2}(x),\:k_\ell(x)=\sqrt{\frac{2}{\pi x}}K_{\ell+1/2}(x).
\]
Only $k_\ell(x)$ is compatible with the boundary condition~(\ref{eqn:bound_infinity}) and goes to zero at infinity, so the solution for $r>R$ is
\begin{equation}\label{eqn:potout}
 f^{\text{out}}(r)= k_\ell(\kappa r).
\end{equation}
In the Coulomb limit~(\ref{eqn:potout}) reduces to $f^{\text{out}}(r)\propto r^{-(\ell+1)}$.

Now let us turn our attention to the plasma region, $r<R$. Here the situation is more complicated since the dielectric function depends on the radial distance from the trap center. Writing the radial function as
\begin{equation}\label{eqn:ansatz_inside}
\tilde f(x,\omega)=x^\ell g(x,\omega),
\end{equation}
leads to the following equation for $g(x,\omega)$,
\begin{align}\label{eqn:inside}
x\frac{\partial}{\partial x} \left[\epsilon(x,\omega) g'(x,\omega)\right]+2(\ell+1)\epsilon(x,\omega)g'(x,\omega) \\
-\left[x-\ell\epsilon'(x,\omega)\right]g(x,\omega)&=0.\nonumber
\end{align}
We now perform another change of variables from $x$ to $z$ via $x^2/x_s^2=z$ with $x_s^2=-2{\Omega}^2 \epsilon_1=2(\Omega_p^2(0)-\Omega^2)$, accompanied by $g(x,\omega)\to\tilde g(z,\Omega)$. Using the explicit result~(\ref{eqn:dielectric}) for the dielectric function, Eq.~(\ref{eqn:inside}) turns into a hypergeometric differential equation for $\tilde g(z,\Omega)$,
\begin{align}
 z(1-z)\tilde g''(z,\Omega)+\left[\ell+3/2 -(\ell+5/2)z\right] \tilde g'(z,\Omega)&\\
-\frac{\ell-{\Omega}^2}{2}\tilde g(z,\Omega)&=0,\nonumber
\end{align}
which has the general solution
\begin{align}\label{eqn:hyper_sol}
\tilde g(z,\Omega)=&A\,_2F_1\left(\frac{\alpha_\ell-\delta_\ell}{2},\frac{\alpha_\ell+\delta_\ell}{2};\alpha_\ell;z\right) +\nonumber \\
&B z^{1-\alpha_\ell}    \,_2F_1\left(\frac{\beta_\ell-\delta_\ell}{2},\frac{\beta_\ell+\delta_\ell}{2};\beta_\ell;z\right)
\end{align}
around $z=0$. Here $A$ and $B$ are arbitrary constants and the parameters of the hypergeometric function $_2F_1$ are
\[
\alpha_\ell=\ell+\frac{3}{2},\:\: \beta_\ell=\frac{1}{2}-\ell,\:\: \delta_\ell=\sqrt{\ell(\ell+1)+\frac{9}{4}+2 {\Omega}^2}.
\]
From Eqs.~(\ref{eqn:ansatz_inside},\ref{eqn:hyper_sol}) we obtain two independent solutions of Eq.~(\ref{eqn:radialx}) inside the plasma,
\begin{align}
\tilde f^{(1)}(x,\omega)&=x^\ell\,_2F_1\left(\frac{\alpha_\ell-\delta_\ell}{2},\frac{\alpha_\ell+\delta_\ell}{2};\alpha_\ell;\frac{x^2}{x_s^2}\right),\nonumber\\
\tilde f^{(2)}(x,\omega)&=x^{-(\ell+1)}\,_2F_1\left(\frac{\beta_\ell-\delta_\ell}{2},\frac{\beta_\ell+\delta_\ell}{2};\beta_\ell;\frac{x^2}{x_s^2}\right).\nonumber
\end{align}
Only $\tilde f^{(1)}(x)$ is finite at the origin and thus constitutes the correct solution for $r\le R$.

Collecting the previous results, the solution inside the plasma is given by
\begin{align}\label{eqn:potin}
 f^{{\text{in}}}(r,\omega)= r^\ell\,_2F_1\left(\frac{\alpha_\ell-\delta_\ell}{2},\frac{\alpha_\ell+\delta_\ell}{2};\alpha_\ell;\frac{\kappa^2 r^2}{x_s^2}\right).
\end{align}
The hypergeometric function describes how the perturbed potential is modified for a Yukawa plasma when compared with the solution for Coulomb interaction, where $f^{{\text{in}}}(r) =r^\ell$.

Since the normal modes explicitly depend on the eigenfrequencies their discussion will be postponed to Sec.~\ref{ssec:normalmodes2}.

\subsection{Eigenfrequencies}\label{ssec:eigenfrequencies}
\subsubsection{Existence of an upper bound for the eigenfrequencies}
Before we explicitly discuss the eigenfrequencies we inspect Eq.~(\ref{eqn:poisson_diel}) more closely. Following Ref.~\cite{dubin2005} we multiply by $\hat \phi_1^*$ and integrate over volume. Using Gauss's theorem we obtain
\begin{equation}
 \int_{\mathbb{R}^3} \left(\kappa^2 |\hat \phi_1|^2 +\epsilon(\kappa r,\omega)|\nabla\hat\phi_1|^2 \right)d\vec r=0.
\end{equation}
The first integral is always positive which implies that there must be a region where $\epsilon(\kappa r,\omega)<0$, i.e. $0<\Omega^2<\text{max}[\Omega_p^2(x)]$, see Eq.~(\ref{eqn:dielectric}). For the density profile considered here the maximum plasma frequency is at the center, $\text{max}[\Omega_p^2(x)]=\Omega_p^2(0)$.

Explicit results for the undetermined eigenfrequencies are now found by requiring Eqs.~(\ref{eqn:bound1}, \ref{eqn:bound2}) to yield non-trivial solutions. Since we used an expansion in spherical harmonics for $\hat\phi_1$, Eq.~(\ref{eqn:bound1}) reduces to the continuity of $f(r,\omega)$ and Eq.~(\ref{eqn:bound2}) requires the continuity of $\epsilon (\kappa r,\omega)f'(r,\omega)$ across the plasma boundary for any given $\ell$. The necessary condition for a non-trivial solution is the vanishing of the determinant,
\begin{widetext}
\begin{equation}\label{eqn:eigenvalues}
\left[\ell\epsilon(\xi)-\xi\frac{k_\ell'(\xi)}{k_\ell(\xi)}\right]\, _2F_1\left(\frac{\alpha_\ell-\delta_\ell}{2},\frac{\alpha_\ell+\delta_\ell}{2};\alpha_\ell;\frac{\xi^2}{x_s^2}\right)+
\epsilon(\xi)\frac{(\ell-{\Omega}^2)}{\alpha_\ell}\frac{\xi^2}{x_s^2}\,_2F_1\left(\frac{\alpha_\ell-\delta_\ell}{2}+1,\frac{\alpha_\ell+\delta_\ell}{2}+1;\alpha_\ell+1;\frac{\xi^2}{x_s^2}\right)=0,
\end{equation}
\end{widetext}
where we used the property
\begin{equation*}
\frac{d\, _2F_1(a,b;c;z)}{dz}=\frac{ab}{c}\, _2F_1(a+1,b+1;c+1;z).
\end{equation*}

Note that Eq.~(\ref{eqn:eigenvalues}) only involves $\Omega^2$ and $\xi$. The eigenfrequency $\omega$ can easily be extracted from $\Omega^2$ by solving $\Omega^2=\omega(\omega+i\nu)$ for $\omega$, which yields the same relation as for the normal modes in the discrete $N$-particle system (damped harmonic oscillator)~\cite{henning09}. In the absence of damping we have $\Omega^2=\omega^2/\omega_0^2$, i.e. $\Omega$ reduces to the eigenfrequency scaled by the trap frequency.
Before we proceed to the solutions of Eq.~(\ref{eqn:eigenvalues}) let us discuss some properties of the hypergeometric series~\cite{wolfram}
\begin{align}\label{eqn:2F1series}
 _2F_1(a,b;c;z)&=\sum_{k=0}^{\infty}\frac{(a)_k (b)_k}{(c)_k}\frac{z^k}{k!}\\
&=1+\frac{ab}{c}\frac{z}{1!}+\frac{a(a+1)\,b(b+1)}{c(c+1)}\frac{z^2}{2!}+\dots,\nonumber
\end{align}
where $(a)_k=\Gamma(a+k)/\Gamma(a)$ denotes the Pochhammer symbol. Its convergence is assured for $|z|<1$ if $c$ is not a negative integer and for $|z|=1$ if $\Re(c-a-b)>0$. In our case the condition $\Re(c-a-b)>0$ is not satisfied as $c-a-b=0$ (or $-1$ for the derivative).

We now apply this to Eqs.~(\ref{eqn:potin},\ref{eqn:eigenvalues}). The convergence condition $|z|=|\kappa^2 r^2/x_s^2|\le |\xi^2/x_s^2|<1$ is closely connected to the plasma frequency at the boundary,
\begin{equation}\label{eqn:omegamax}
 {\Omega}_p^2(\xi)=3+\frac{\xi^2}{1+\xi},\hspace{0.6cm} \xi>0,
\end{equation}
since it is fulfilled for all $\Omega^2<\Omega_p^2(\xi)$.

To better understand the nature of this maximum frequency we recall the Coulomb limit. In this case the density is $r$ independent and there exists a unique plasma frequency, $\Omega_p=\sqrt{3}$, defined by Eqs.~(\ref{eqn:plasmafreq},\ref{eqn:density0}), at which the dielectric function~(\ref{eqn:dielectric}) vanishes. This is in contrast to Yukawa interaction due to the inhomogeneous density profile. However, the frequency~(\ref{eqn:omegamax}) is the plasma frequency at $r=R$ ($x=\xi$), i.e. it is precisely the one for which the dielectric function vanishes at the plasma boundary, $\epsilon(\xi,\Omega_p(\xi))=0$. 

For any $\Omega^2$ in the interval $\Omega_p^2(0)>\Omega^2>\Omega_p^2(\xi)$ there exists a point $x_s$ inside the plasma at which $\Omega^2=\Omega_p^2(x_s)$ and consequently $\epsilon(x_s,\Omega)=0$. At this point the local plasma frequency is in resonance with the mode frequency and the differential equation~(\ref{eqn:radialx}) has a singular point. In these cases the solutions of~(\ref{eqn:radialx}) can be singular and are associated with a continuous spectrum and, possibly, damped quasi-modes, see e.g.~\cite{barston1964,dubin2005}. The appropriate approach for uncovering the quasi-modes is a Laplace transform of Eqs.~(\ref{eqn:perturbed}). Here we restrict ourselves to regular normal mode solutions with $\Omega^2<\Omega_p^2(\xi)$.

\subsubsection{Coulomb limit and frequency degeneracy lifting of bulk modes in a Yukawa plasma}
Having obtained these general properties we now explicitly determine the eigenfrequencies. Let us first consider the limit $\xi\ll 1$. Performing a series expansion of Eq.~(\ref{eqn:eigenvalues}) for $\ell\neq 0$ (details can be found in Appendix~\ref{sec:AppendixB}) we find
\begin{equation}\label{eqn:smallxiexp}
\Omega_\ell^2(\xi)\simeq \frac{3\ell}{2\ell+1}+\left(\frac{4\ell^3+6\ell^2-10\ell}{8\ell^3+12\ell^2-2\ell-3}\right)\xi^2+\dots\, .
\end{equation}
For Coulomb interaction ($\xi = 0$) we get ${\Omega}_\ell^2= 3\ell/(2\ell+1)$. This is the well known result for surface oscillations of a homogeneous plasma sphere~\cite{dubin_corr96}. Furthermore it is easily verified that $\Omega_1=1$ is a solution of Eq.~(\ref{eqn:eigenvalues}) for any $\xi$. This mode describes the center of mass oscillation (dipole or sloshing mode) and is independent of the particle number and the screening parameter~\cite{bonitz2007}.

Another solution in the Coulomb limit is given by the Coulomb plasma frequency $\Omega_p$. This can easily be seen from Eq.~(\ref{eqn:poisson_diel}) with $\kappa=0$. Since the dielectric function inside the plasma vanishes for $\Omega=\Omega_p$, Eq.~(\ref{eqn:poisson_diel}) is satisfied for any $\hat{\phi}_1^\text{in}$ that satisfies the boundary conditions~(\ref{eqn:boundarycond}).  This implies a high degeneracy since these requirements can be met by an infinite number of modes with arbitrary $\ell,m$.

For Yukawa interaction with a finite $\xi$ this degeneracy is lifted and a series expansion yields (see Appendix~\ref{sec:AppendixB})
\begin{equation}\label{eqn:smallxiexp2}
 \Omega_\ell^2(\xi)\simeq 3+c_\ell \xi^2+\dots,\hspace{1cm}\xi\ll 1,\\
\end{equation}
where $c_0\approx 0.85031$, $c_1\approx 0.98624$ and $c_2\approx 0.99992$. From this series of expansion coefficients we see that the frequency approaches $\Omega_p^2(\xi)\simeq 3+\xi^2+\dots$ as $\ell$ increases.

For $\Omega^2=\Omega_p^2(\xi)$ the hypergeometric series does not converge in general at $|z|=1$. However, a closer inspection of the series representation~(\ref{eqn:2F1series}) reveals that a well behaved solution can be obtained if $a$ or $b$ is a negative integer $-(n-1)$. In these cases the series is simply a polynomial of order $n-1$, where $n\in \mathbb{N}^+$. Since $b=(\alpha_\ell+\delta_\ell)/2>0$ we require $a=(\alpha_\ell-\delta_\ell)/2=-(n-1)$. Solving this equation for $\Omega^2$ yields
\begin{align}\label{eqn:omegamaxnl}
 \Omega^2_{n \ell}(\xi_{n \ell}^\text{crit})=(2n+1)(n-1)+(2n-1)\ell.
\end{align}
Keeping in mind that at this point $\Omega_p^2(\xi)=3+\xi^2/(1+\xi)$ we find
\begin{align}\label{eqn:xicrit}
 \xi_{n \ell}^\text{crit}&=\frac{1}{2}\left[ \zeta_{n \ell}+\sqrt{\zeta_{n \ell}(\zeta_{n \ell}+4)}\right],\nonumber\\
\zeta_{n \ell}&= (2n-1)(n+\ell)-4,
\end{align}
where $\ell\ge 3$ for $n=1$. Even though this choice guarantees a well behaved solution, it can be shown that Eqs.~(\ref{eqn:omegamaxnl},\ref{eqn:xicrit}) do not solve the eigenvalue equation~(\ref{eqn:eigenvalues}). Nevertheless, a numerical evaluation shows that solutions do exist for $\xi>\xi_{n \ell}^\text{crit}$ which closely approach $\Omega_p(\xi)$ as $\xi\to\xi_{n \ell}^\text{crit}$. This issue will be dealt with in more detail in Sec.~\ref{ssec:normalmodes2}.

\subsubsection{Macroscopic/strong screening limit, $\xi\gg 1$}
Let us now discuss the limit $\xi\to \infty$. Using $\Omega^2<\Omega_p^2(\xi)$ one can show that $|\xi^2/x_s^2|\to 1$. As before, the convergence problem at this point can be circumvented by choosing the parameters of the hypergeometric function such that its series terminates at a finite order. Requiring $a=(\alpha_\ell-\delta_\ell)/2=-n$ ($n\in \mathbb N$) and solving the equation for ${\Omega}^2$ yields the solution for the eigenfrequencies in the limit $\xi\to\infty$,
\begin{equation}\label{eqn:freqlimit}
 \lim_{\xi\to\infty}{\Omega}_{n \ell}^2(\xi)\equiv \Omega_{n \ell,\infty}^2 =2n^2+(2\ell+3)n+\ell.
\end{equation}
The reason we chose the same index $n$ as in the previous case will become clear shortly. 

It is shown in Appendix~\ref{sec:AppendixB} that~(\ref{eqn:freqlimit}) actually solves Eq.~(\ref{eqn:eigenvalues}). Further, the lowest order correction for finite $\xi$ is found as
\begin{equation}\label{eqn:freqlimitTaylor}
 {\Omega}_{n \ell}^2(\xi)\simeq  \Omega_{n \ell,\infty}^2 -\frac{d_{n \ell}}{\xi^2}+\dots, \hspace{1cm}\xi\gg 1,
\end{equation}
where the coefficients are given by
\begin{align}\label{eqn:coefflargexi}
 d_{n \ell}=&\, (2n+\ell+3/2)[(4n^3+12n^2+3n-9)n\nonumber \\
&+ 2\ell(4n^3+8n^2+2\ell n(n+1)+\ell-1)].
\end{align}
\begin{figure}
\includegraphics[width=0.48\textwidth]{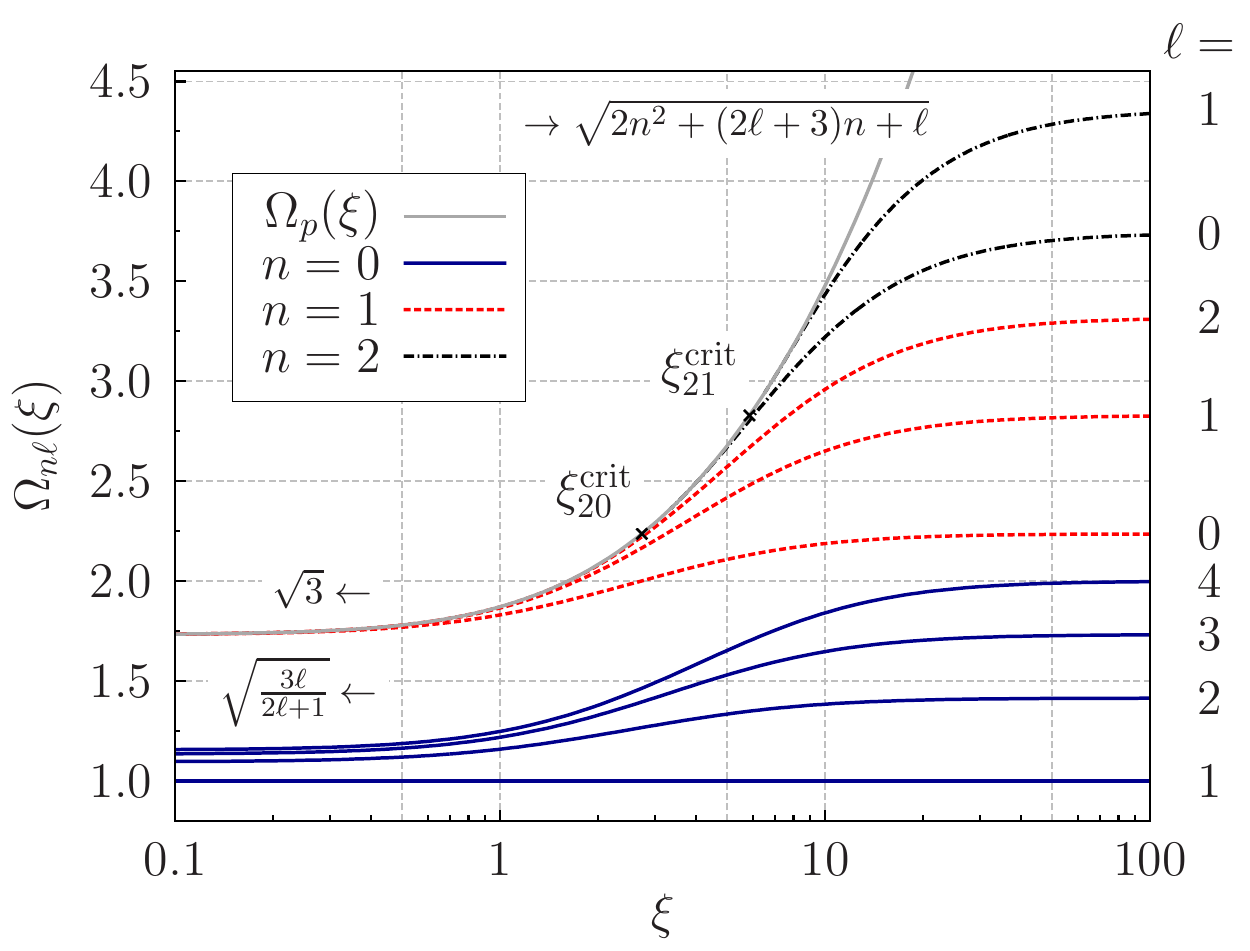}
\caption{Eigenfrequencies and their dependence on $\xi$ for various modes $(n ,\ell)$. Also shown (by the arrows) are the limits for $\xi=0$ and $\xi\to\infty$. The crosses denote the parameters at which new modes appear to the right of $\xi^{\text{crit}}_{n\ell}$.}\label{fig:eigenfrequencies}
\end{figure}

\subsubsection{Eigenfrequencies for arbitrary $\xi$}
For arbitrary values of $\xi$ we solved Eq.~(\ref{eqn:eigenvalues}) numerically. The results are shown in Fig.~\ref{fig:eigenfrequencies}. In the weak screening limit, $\xi\ll 1$, the known Coulomb limit is recovered, where $\Omega_\ell^2= 3\ell/(2\ell+1)$ (surface modes) or $\Omega_p^2=3$ (bulk modes)~\cite{dubin_corr96}. Except for the center of mass mode all mode frequencies increase with $\xi$ and saturate in the limit $\xi\to\infty$. We find numerically that the eigenmodes with mode number $n\ge 1$ ($\ell\ge 3$ for $n=1$) at $\xi=\infty$ approach $\Omega_p(\xi)$ as $\xi$ is decreased and cease to exist for $\xi\le\xi_{n \ell}^\text{crit}$. Thus, the chosen indices are the same and the modes evolve continuously from $\xi_{n \ell}^\text{crit}$ to $\xi=\infty$.

Let us briefly summarize the findings of this section. The main result is that screening lifts the degeneracy of the bulk modes of the Coulomb system and the appearance of a new mode number $n$. For a given value of $\xi$ the number of allowed modes is restricted. Modes with $n=0$ ($\ell\ge 1$) and $n=1$ with $\ell=0,1,2$ exist for all $\xi\ge 0$ whereas our numerical solution of Eq.~(\ref{eqn:eigenvalues}) indicates that all other modes exist only for $\xi>\xi_{n \ell}^\text{crit}$. The $(n ,\ell)=(1,3)$ mode is a special case with $\xi_{13}^\text{crit}=0$. 

Having found the eigenfrequencies we can now come back to the discussion of the shape of the normal modes.

\subsection{Explicit results for the normal modes}\label{ssec:normalmodes2}
\subsubsection{Coulomb limit}
The eigenmodes in the Coulomb limit are well known, see e.g.~\cite{dubin_md96}. The surface modes with $\Omega_\ell^2=3\ell/(2\ell+1)$ are given by $\hat{\phi}_1^\text{out}\sim r^{-(\ell+1)} Y_\ell^m(\theta,\varphi)$ and $\hat{\phi}_1^\text{in}\sim r^\ell Y_\ell^m(\theta,\varphi)$, cf. Eqs.~(\ref{eqn:potout},\ref{eqn:potin}), while the bulk modes oscillate at the plasma frequency $\Omega_p=\sqrt{3}$. The potential eigenfunctions in the latter case are undefined inside the plasma, which can easily be seen from Eq.~(\ref{eqn:poisson_diel}). Since the dielectric function in the Coulomb limit is constant for $r\le R$, Eq.~(\ref{eqn:poisson_diel}) is satisfied for any $\hat{\phi}_1^\text{in}$ if $\Omega=\Omega_p$. The potential perturbation is only restricted by the boundary conditions~(\ref{eqn:boundarycond}). It follows from Eqs.~(\ref{eqn:boundarycond}) and $\Delta \hat{\phi}_1^\text{out}=0$ that $\hat{\phi}_1^\text{out}=0$~\cite{dubin_md96}. This further implies that $f^\text{in}(R)=0$, see Eq.~(\ref{eqn:bound1}).

In the following we will discuss the eigenmodes for Yukawa interaction and point out the similarities and differences when compared with the Coulomb limit.

\subsubsection{General remarks for $\xi>0$}
For $\xi>0$ the radial eigenfunctions~(\ref{eqn:potin}) inside the plasma explicitly depend on $\Omega^2$, which was shown to have several solutions for a given $\ell$. Thus, in addition to the angular mode numbers $m$ and $\ell$, there is a radial mode number $n$ which determines the structure of the radial eigenfunction. The radial function is given by Eq.~(\ref{eqn:potin}) and the corresponding eigenfrequency is determined by Eq.~(\ref{eqn:eigenvalues}).

\subsubsection{Eigenmodes for Yukawa interaction}
Let us begin with the limit $\xi\to\infty$. It was already shown in Sec.~\ref{ssec:eigenfrequencies} that in this case the hypergeometric function reduces to a polynomial of order $n$ in $z=r^2/R^2$. From Eqs.~(\ref{eqn:potin},\ref{eqn:freqlimit}) we thus get the radial eigenfunctions
\begin{align}\label{eqn:modelimit}
f^\text{in}(r,\omega_{n \ell}^\infty)&=  r^\ell\sum_{k=0}^n \frac{(-n)_k}{k!} \frac{(\alpha_\ell+n)_k}{(\alpha_\ell)_k}\left(\frac{r}{R}\right)^{2k}\\
&\propto  r^\ell P_n^{(\ell+1/2,0)}\left(1-\frac{2 r^2}{R^2}\right),\nonumber
\end{align}
where the $P_n^{(\ell+1/2,0)}(x)$ are Jacobi polynomials. The $n=0$ surface modes are particularly simple. Here, the sum in Eq.~(\ref{eqn:modelimit}) is just a constant and the eigenmodes inside the plasma are the same as in the Coulomb limit. There are no radial nodes. For finite $\xi$ we find that the solutions differ only slightly from the results at $\xi=0$ or $\infty$. These modes are the natural generalization of the surface modes to Yukawa interaction and there are no qualitative changes compared to a Coulomb system. In particular, the dipole modes with $(n, \ell)=(0,1)$ describe the three center of mass oscillations and have eigenfunctions $\hat{\phi}^{\text{in}}_1\sim r\,Y_1^m(\theta,\varphi)$, independent of $\xi$. 

\begin{figure}
\includegraphics[width=0.47\textwidth]{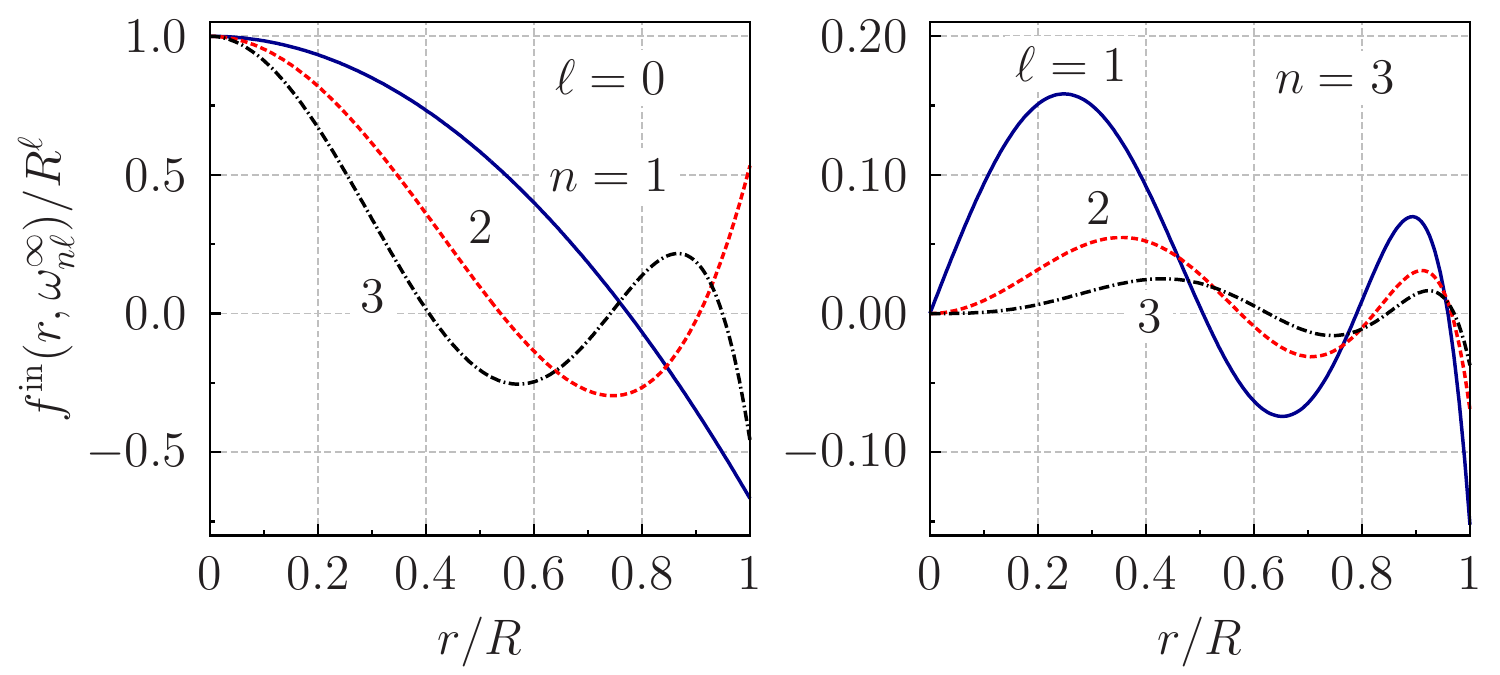}
\caption{Radial eigenfunctions in the limit $\xi\to \infty$ for various modes as indicated in the figure.}\label{fig:infmodes}
\end{figure}

The origin of the modes with $n>0$ can be traced back to the bulk modes of the Coulomb system, see Fig.~\ref{fig:eigenfrequencies}. While for $\xi=0$ the eigenmodes are not entirely specified, for $\xi>0$ their form is determined by Eq.~(\ref{eqn:potin}), together with~(\ref{eqn:eigenvalues}) for the eigenfrequency. In the limit $\xi\to\infty$ the potential perturbations are given by Eq.~(\ref{eqn:modelimit}), see Fig.~\ref{fig:infmodes}. $n$ is the number of radial nodes. With increasing $\ell$ the nodes and extrema are shifted towards the cluster boundary.

\begin{figure}
\includegraphics[width=0.47\textwidth]{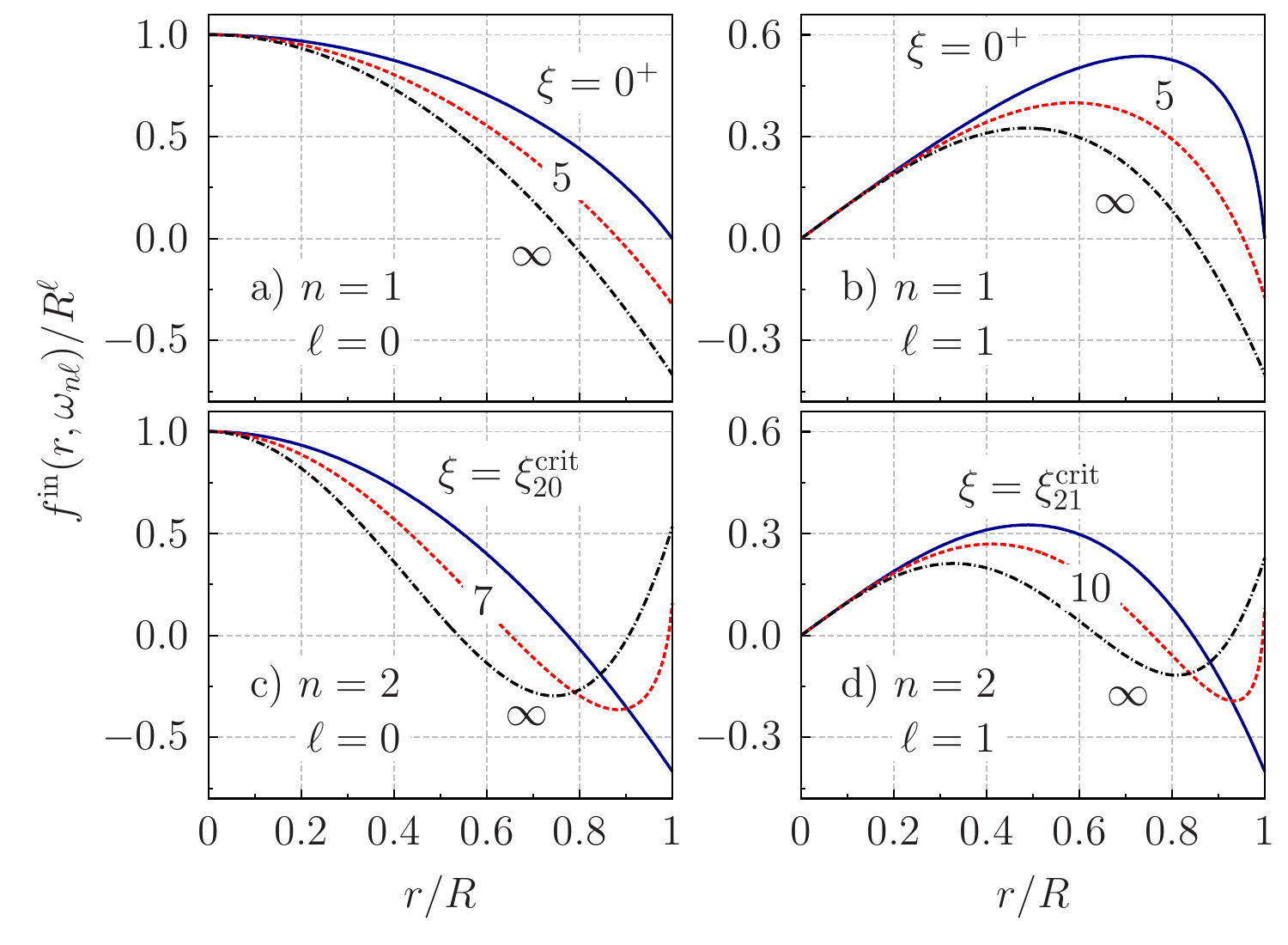}
\caption{Radial eigenfunctions $f_{n \ell}(r)$ for various $\xi$ and mode numbers $(n ,\ell)$.}\label{fig:n1n2modes}
\end{figure}

The behavior for finite $\xi$ is shown in Fig.~\ref{fig:n1n2modes}. The $n=1$ modes with $\ell=0,1,2$ extend up to $\xi=0$, while all other modes with $n\ge 1$ exist only for $\xi>\xi_{n \ell}^\text{crit}$.

Let us discuss the former case first. As we move from the $\xi=\infty$ limit towards $\xi=0$, the single node approaches $r=R$, see Fig.~\ref{fig:n1n2modes}a,b. The radial eigenfunctions at $\xi=0^+$ read
\begin{align*}
 f^\text{in}(r,\omega_{1\ell}^0)=r^\ell\, _2F_1\left(\frac{\alpha_\ell-\delta_\ell}{2},\frac{\alpha_\ell+\delta_\ell}{2};\alpha_\ell;\frac{(r/R)^2}{3-2\bar c_\ell}\right),
\end{align*}
where $\ell=0,1,2$ and $\delta_\ell$ must be evaluated at $\Omega_p^2=3$. Compare also with the series expansion performed in Appendix~\ref{sec:AppendixB}. This form is in accordance with $f^\text{in}(R)=0$ for $\xi=0$ (the coefficients $\bar c_\ell$ were chosen to ensure this). Thus, these modes also exist in a Coulomb system, where they are among the bulk modes with $\Omega=\Omega_p$. The main difference between Coulomb and Yukawa interaction at this point is that in the former case the mode form is not specified by Eq.~(\ref{eqn:poisson_diel}). For $\xi>0$ this equation constitutes an additional restriction on the mode form and selects the eligible modes for $\xi=0^+$ from the large number of modes at $\xi=0$.

In the case of all other modes with $n\ge 1$ the outermost node comes arbitrarily close to $r=R$ as $\xi$ decreases, but disappears at $\xi_{n \ell}^\text{crit}$, see Fig.~\ref{fig:n1n2modes}c,d. This goes along with a very strong increase of $\partial f^\text{in}(r)/\partial r$ at the boundary, which makes the numerical solution of Eq.~(\ref{eqn:eigenvalues}) increasingly difficult. If $\Omega=\Omega_p(\xi_{n \ell}^\text{crit})$ were a proper solution, the boundary conditions~(\ref{eqn:boundarycond}) and Eq.~(\ref{eqn:radialout}) would require $f^\text{out}=0$, and hence $f^\text{in}(R)=0$, just like for $\xi=0$ and $\Omega=\Omega_p$. However, the potential inside the plasma is not undefined but determined by the solutions of Eq.~(\ref{eqn:poisson_diel}). Analogously to the case $\xi=\infty$, Eq.~(\ref{eqn:potin}) reduces to a polynomial at $\Omega_p(\xi_{n \ell}^\text{crit})$ but cannot satisfy the boundary conditions since $f^\text{in}(R)\ne 0$. Our numerical solutions of Eq.~(\ref{eqn:eigenvalues}) indicate that these modes exist only for $\xi>\xi_{n \ell}^\text{crit}$.

\subsubsection{Breathing mode}
\begin{figure}
\includegraphics[width=0.4\textwidth]{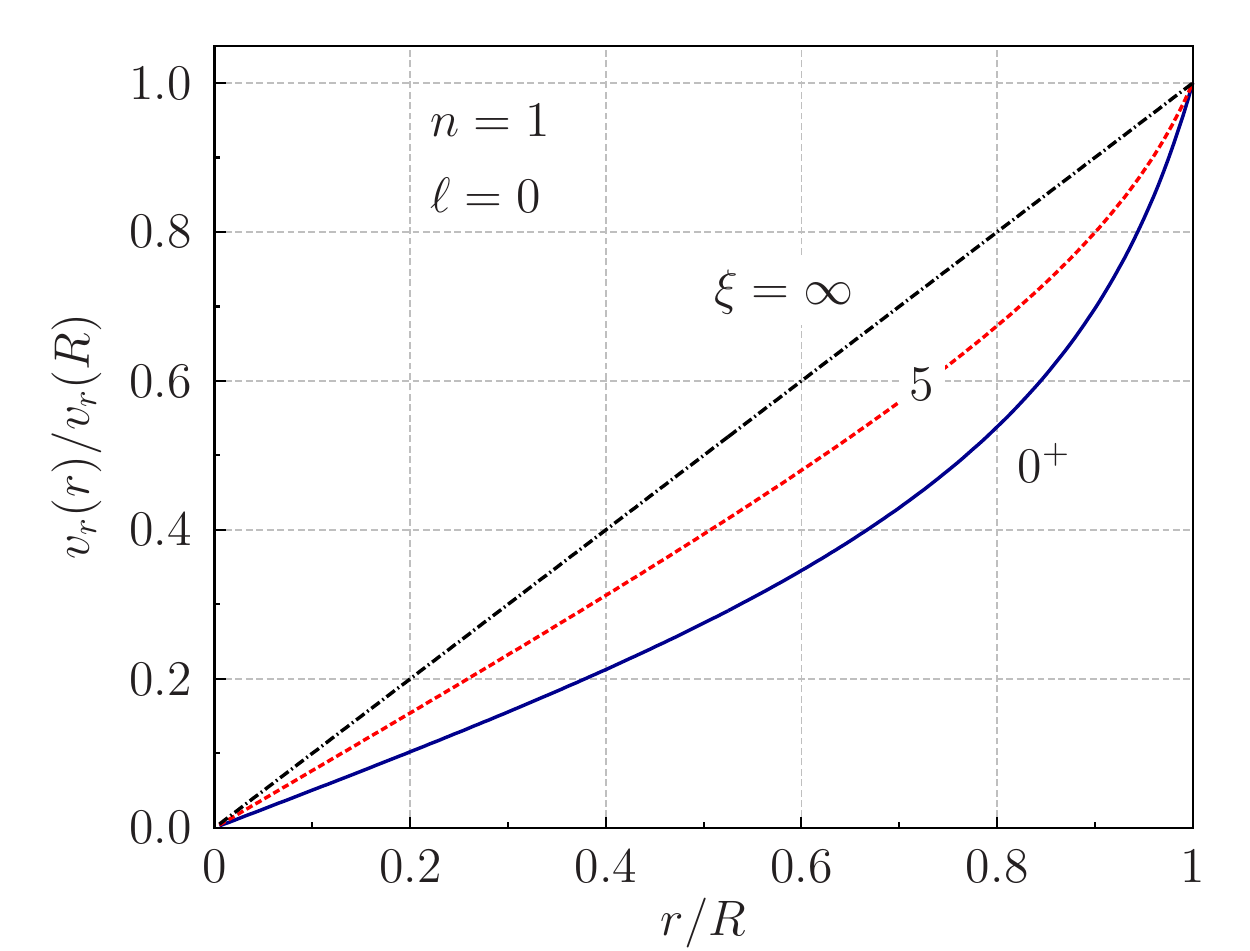}
\caption{Radial fluid velocity for the breathing mode.}\label{fig:breathingmode}
\end{figure}

The lowest monopole mode has the index $(n ,\ell)=(1,0)$ and is worth a more detailed discussion. For $\xi\to\infty$ the fluid velocity satisfies $\hat{\vec v}_1^{\text{in}}\sim r\hat{e}_r$ since $\hat{\phi}^{\text{in}}_1$ is quadratic in $r$, see Eq.~(\ref{eqn:vel1}). This corresponds to a uniform breathing oscillation of the plasma. In the Coulomb limit this mode is among the bulk modes. It was shown in Ref.~\cite{henning08} that a system of $N$ harmonically confined particles with Yukawa interaction does not support a universal uniform breathing mode. We find a similar behavior in our fluid approach since the uniformity condition is only fulfilled for $\xi\to\infty$. For finite $\xi$ the uniformity condition is not satisfied, see Fig.~\ref{fig:breathingmode}. Further, it is surprising that it is not recovered as $\xi\to 0$. However, this is not contradictory since this solution is just one among many in the Coulomb limit. Here, the breathing mode is given by $f^\text{in}(r)\sim (r^2 -R^2)$, which also satisfies the boundary condition $f^\text{in}(R)=0$~\cite{dubin_md96}.

The aforementioned breathing mode has been studied by Sheridan~\cite{sheridan2006} under the assumption of a homogeneous density and a uniform displacement. We find that his analytical result for $\Omega^2(\xi)$ agrees with our numerical results to within $0.2\%$ for any $\xi$. Note, however, that the equation $\xi$ is determined from in~\cite{sheridan2006} is different from Eq.~(\ref{eqn:radius}), since $\xi$ used in~\cite{sheridan2006} involves the radius for a \textit{homogeneous} sphere. This mode was found to be the dominant excitation during spherical crystal formation after a rapid temperature quench~\cite{kaehlert2010}.

\section{Conclusion}\label{sec:conclusion}
Summarizing our results, we have investigated the ground state and the normal modes of a harmonically confined Yukawa plasma in a fluid approach. The results of~\cite{Christian2006} were extended to a time-dependent theory and simple formulas for the plasma radius, total energy and density moments were obtained. The density moments could be used to infer the dimensionless plasma parameter $\xi=\kappa R$, the ratio of the plasma radius to the screening length, from experimental data. For typical dusty plasma experiments with $\kappa a \approx 0.5\dots 1$ and $N\approx 50\dots 1000$ we expect $\xi\approx 1.5\dots 6$~\cite{Bonitz2006}.

Further, the fluid equations were linearized and solved for the normal modes. Compared with previous results for Coulomb interaction, we found a new class of modes with a radial mode number $n$ that determines the number of radial nodes in the potential eigenfunctions. The eigenfrequencies were found to depend only on $\xi$. The degeneracy of the bulk modes in the Coulomb limit was shown to be lifted for Yukawa interaction and series expansions for the eigenfrequencies were derived for limiting cases. For experimentally relevant parameters they must be determined numerically, though.

The fluid theory should be applicable for large clusters with $N\gtrsim 100$. In~\cite{Christian2006} it was shown that the agreement between the continuum theory and the exact $N$-particle ground state was good at low screening, but deviations increased for larger $\kappa a$. Applying the local density approximation, the authors were able reduce the deviations for strong screening~\cite{Christian2007}. The same behavior was observed in~\cite{kaehlert2010}, where the frequency of the monopole mode excited in a Langevin dynamics simulation was compared with the theory of~\cite{sheridan2006}. Thus, similar behavior is expected for the results presented here. Additionally, one has to bear in mind that the fluid equations correspond to a mean-field description and neglect correlation effects. They are thus not able to describe crystallization and shell structure formation. The same applies to the local density approximation used in~\cite{Christian2007}. This question was recently analyzed in~\cite{wrighton2009, wrighton2010}. Only a comparison with first principle simulation data will show the true applicability limits of the fluid approach. This analysis is subject of ongoing work.

\begin{acknowledgments}
This work is supported by the Deutsche Forschungsgemeinschaft via SFB-TR24.
\end{acknowledgments}

\appendix
\section{Calculation of $\phi_0(r)$}\label{sec:AppendixA}
Carrying out the angle integration in Eq.~(\ref{eqn:potential0}) we obtain~\cite{Christian2006}
\begin{align}
\phi_0(r)=2\pi \frac{q^2}{\kappa r} \int_0^R dr'\, n_0(r')\,r'\left[e^{-\kappa|r-r'|}-e^{-\kappa(r+r')} \right].\nonumber
\end{align}
The density involves a constant term and a term $\sim r^2$. Thus we consider the following integrals for $n=1,\,3$,
\begin{align}
\mathcal{I}^{\gtrless}_n=\int_0^R dr'\, (r')^n\left[e^{-\kappa|r-r'|}-e^{-\kappa(r+r')} \right].\nonumber
\end{align}
If $r>R$ we have $|r-r'|=r-r'$ for the entire integration and
\begin{align}
\mathcal{I}^>_n&=e^{-\kappa r}\int_0^R dr'\, (r')^n\left[e^{\kappa r'}-e^{-\kappa r'} \right]=2\,\frac{e^{-\kappa r}}{\kappa^4}\nonumber\\
&\times\begin{cases}
 [\xi \cosh(\xi)-\sinh(\xi)]\kappa^2, &n=1, \\
 \xi(\xi^2+6)\cosh(\xi)-3(\xi^2+2)\sinh(\xi), &n=3.
\end{cases}\nonumber
\end{align}
For $r\le R$ the integrals must be solved independently in regions where $r'<r$ and $r'>r$,
\begin{align}
\mathcal{I}^<_n&=\mathcal{I}_n^>(\xi\to \kappa r)+\int_r^{R} dr'(r')^n\left[e^{-\kappa(r'-r)} -e^{-\kappa(r+r')}\right]\nonumber\\
&=\mathcal{I}_n^>(\xi\to \kappa r)+2\sinh(\kappa r)\int_r^R dr'\,(r')^n e^{-\kappa r'}.\nonumber
\end{align}
The remaining integral can easily be solved. Collecting the results and using the explicit result for $n_0(r)$ [Eq.~(\ref{eqn:density0})] we obtain Eq.~(\ref{eqn:potential0sol}). Alternatively, the potential inside the plasma may be directly calculated from Eqs.~(7,10) of~\cite{Christian2006}.

\section{Series expansion for $\Omega^2(\xi)$}\label{sec:AppendixB}
The difficulty in finding an expansion for $\Omega^2(\xi)$ arises from the hypergeometric function $_2F_1(a,b;c;z)$ since $a,b$ and $z$ are all functions of $\xi$. In the following we will separately discuss two different limits.
\subsection{Coulomb limit, $\xi\ll 1$}
Expanding the squared eigenfrequency as $\Omega^2(\xi)\approx \bar a +\bar b\xi+\bar c\xi^2$ we find
\begin{align*}
\epsilon(\xi)&\approx 1-\frac{3}{\bar a}+\left(\frac{3\bar b}{\bar a^2}\right)\xi+\left(\frac{3\bar a\bar c-3\bar b^2-\bar a^2}{\bar a^3}\right)\xi^2.
\end{align*}
Further, we obtain $x_s^2\approx 6-2\bar a$ and $-\xi \,k_\ell'(\xi)/k_\ell(\xi)\approx (\ell+1)+\xi^2/(2\ell-1)$ for $\ell\ne 0$ or $-\xi \,k_0'(\xi)/k_0(\xi)= 1+\xi$. The series expansion for the hypergeometric function, Eq.~(\ref{eqn:2F1series}), then yields, for $\bar a\ne 3$,
\begin{align*}
 _2F_1\left(\frac{\alpha_\ell-\delta_\ell}{2},\frac{\alpha_\ell+\delta_\ell}{2};\alpha_\ell;\frac{\xi^2}{x_s^2}\right)\approx 1+\left(\frac{\bar a-\ell}{\bar a-3}\right)\frac{\xi^2}{6+4\ell}.
\end{align*}
It is sufficient to approximate the other hypergeometric function in Eq.~(\ref{eqn:eigenvalues}) by $1$. Comparing terms of order $\xi^0,\xi^1,\xi^2$, we find the coefficients $\bar a,\bar b,\bar c$ as given in Eq.~(\ref{eqn:smallxiexp}).

If $\bar a=3$ we must keep in mind that $\Omega^2(\xi)<\Omega_p^2(\xi)$. The ansatz $\Omega^2(\xi)\approx 3+\bar b\xi$ yields $\xi^2/x_s^2\approx -\xi/(2\bar b)$. However, this choice turns out to be inadequate since it leaves a constant term in Eq.~(\ref{eqn:eigenvalues}). Next, we use $\Omega^2(\xi)\approx 3+\bar c\xi^2$, which results in $\xi^2/x_s^2\approx 1/(3-2\bar c)$. Then, the lowest order term in Eq.~(\ref{eqn:eigenvalues}) vanishes if
\begin{align*}
\left. _2F_1\left(\frac{\alpha_\ell-\delta_\ell}{2},\frac{\alpha_\ell+\delta_\ell}{2};\alpha_\ell;\frac{1}{3-2\bar c}\right)\right|_{\Omega^2=3}=0.
\end{align*}
This equation can be solved numerically for $\bar c$ if $\ell=0,1,2$ and yields the coefficients given below Eq.~(\ref{eqn:smallxiexp2}). For $\ell\ge 3$ we find no solution. The coefficients are in accordance with the condition $\Omega^2(\xi)<\Omega_p^2(\xi)\approx 3+\xi^2$ as they satisfy $\bar c<1$.

\subsection{Macroscopic/strong screening limit, $\xi\gg 1$}
In this limit we seek an expansion in terms of $y=\xi^{-1}\ll 1$. The ansatz $\Omega^2(y)\approx \bar a+\bar by+\bar cy^2$ yields
\begin{align*}
\epsilon(y^{-1})&\approx -\frac{1}{\bar a y}+\frac{\bar a(\bar a-2)+\bar b}{\bar a^2},\:\; \frac{y^{-2}}{x_s^2}\approx 1-2y,
\end{align*}
and $-k_\ell'(y^{-1})/k_\ell(y^{-1})/y\approx y^{-1}+1$. The expansion for the hypergeometric function is found from~\cite{wolfram}
\begin{widetext}
\begin{align*}
 _2F_1(a,b;a+b-m;z)&=\frac{(m-1)!\,\Gamma(a+b-m)}{\Gamma(a)\Gamma(b)}(1-z)^{-m}\sum_{k=0}^{m-1}\frac{(a-m)_k (b-m)_k(1-z)^k}{k!(1-m)_k} +\frac{(-1)^m\Gamma(a+b-m)}{\Gamma(a-m)\Gamma(b-m)}\\
 &\times\sum_{k=0}^\infty \frac{(a)_k (b)_k}{k!(k+m)!}( -\ln(1-z) +\psi(k+1)+\psi(k+m+1)-\psi(a+k)-\psi(b+k))(1-z)^k ,
\end{align*}
\end{widetext}
valid for $|z-1|<1$ and $m\in \mathbb N^+$. $\psi(x)$ denotes the Digamma function. The expansion for $m=0$ is the same, but without the first (finite) sum.

The following calculation is based on $\bar a=\Omega^2_{n \ell,\infty}$, see Eq.~(\ref{eqn:freqlimit}), and will show that this choice solves Eq.~(\ref{eqn:eigenvalues}). For the first hypergeometric function in Eq.~(\ref{eqn:eigenvalues}) we have $m=0$, $a\approx -n-q(y)$ and $b\approx n+\alpha_\ell+q(y)$, where $q(y)\ll 1$. The main contribution in the sum arises from $\psi(a+k)\approx \psi(-n+k-q)\approx q^{-1}$ (for $k\le n$), which yields a constant term when combined with the prefactor $\Gamma^{-1}(a)\approx \Gamma^{-1}(-n-q)\approx (-1)^{n+1} n!\, q$. The lowest order term is then found as
\begin{align*}
 _2F_1(&-n-q,n+\alpha_\ell+q;\alpha_\ell;1-2y)\approx \frac{(-1)^n}{(\alpha_\ell)_n}n!.
\end{align*}
Similarly, we obtain for the hypergeometric function with shifted parameters and $n=1,2,\dots$,
\begin{align*}
 _2F_1(&-n+1-q,n+1+\alpha_\ell+q;\alpha_\ell+1;1-2y)\approx \\
 &\alpha_\ell \frac{(-1)^n}{(\alpha_\ell)_n}n!\left[ \frac{\bar b}{2\chi n(n+\alpha_\ell)}-1 \right],\:\:\chi=4n+2\alpha_\ell,
\end{align*}
where we used the above expansion for $m=1$ and $q(y)\approx \bar by/\chi$. In the following the notation $_2F_1[+1]$ ($_2F_1[+0]$) will be used to denote the (un)shifted hypergeometric function. Comparing terms $\mathcal O(y^{-1})$ we find that Eq.~(\ref{eqn:eigenvalues}) is satisfied if $\bar b=0$. For $n=0$ the previous equation is not valid and $_2F_1[+1]\approx (2y)^{-1}$. In this case the leading order term in Eq.~(\ref{eqn:eigenvalues}) is $\mathcal O(y^{-2})$. It vanishes due to its prefactor $\sim (\ell-\bar a)$, since $\bar a=\ell$ for $n=0$. The $\mathcal O(y^{-1})$ term vanishes for $\bar b=0$.

In order to calculate the lowest order correction to $\Omega^2(y)$ for $n=1,2,\dots$ we need to evaluate the hypergeometric function up to first order in $y$. We find, using $q(y)\approx \bar cy^2/\chi$,
\begin{align*}
 &_2F_1[+0]\approx \frac{(-1)^n}{(\alpha_\ell)_n}n!\left[ 1-2n(n+\alpha_\ell)y\right],\:\:_2F_1[+1]\approx \alpha_\ell\\
  &\frac{(-1)^n}{(\alpha_\ell)_n}n!\left[ -1+\left( (n-1)(n+\alpha_\ell+1)+\frac{\bar c}{2\chi n(n+\alpha_\ell)} \right)y\right].
\end{align*}
The coefficient $\bar c$ can now be determined by choosing it such that terms $\mathcal O(y^0)$ vanish, yielding Eqs.~(\ref{eqn:freqlimitTaylor},\ref{eqn:coefflargexi}) for $n>0$. If $n=0$ the same procedure leads to $\bar c=-2\ell(\ell-1)(\ell+3/2)$, i.e. Eq.~(\ref{eqn:coefflargexi}) also holds for $n=0$ (here $\ell\ne 0$).


\end{document}